\documentclass[twocolumn]{article}
\usepackage[english]{babel}
\usepackage{graphicx,color,authblk,amsmath}
\usepackage{multirow,tabularx}
\usepackage{authblk}
\usepackage{blindtext}
\usepackage[switch]{lineno}
\begin{document}

\title{Tunable drag drop via flow-induced snap-through in origami}
\author[1*]{Rishabh Nain}
\author[1,2]{Tom Marzin}
\author[1]{Sophie Ramananarivo}

\affil[1]{Laboratoire d'Hydrodynamique, CNRS, École polytechnique, Institut Polytechnique de Paris, 91120 Palaiseau, France}
\affil[2]{Chemical and Biological Engineering, Princeton University,
NJ 08544, United States}
\affil[*]{\textit {rishabh.nain@ladhyx.polytechnique.fr}}

\date{}
\twocolumn[
\maketitle
\begin{@twocolumnfalse}
\vspace{-1cm}
\begin{abstract}

We leverage the snap-through response of a bistable origami mechanism to induce a discontinuous evolution of drag with flow speed. The transition between equilibrium states is passively actuated by airflow, and we demonstrate that large shape reconfiguration over a small increment of flow velocity leads to a pronounced and sudden drop in drag. Moreover, we show that systematically varying the geometrical and mechanical properties of the origami unit enables the tuning of this drag discontinuity and the critical speed and loading at which it occurs. Experimental results are supported by a theoretical aero-elastic model, which further guides inverse design to identify the combination of structural origami parameters for targeted drag collapse. This approach sheds light on harnessing origami-inspired mechanisms for efficient passive drag control in a fluid environment, applicable for load alleviation or situations requiring swift transitions in aerodynamic performances. 

\end{abstract}
\vspace{0.5cm}
\end{@twocolumnfalse}
]

The drag force acting on an object typically increases with the speed of the surrounding fluid flow, as the pressure and shear stress exerted on its surface grow. For flexible bodies, this increase can be partially offset by the deformations induced by fluid forces, as it produces a more streamlined shape and reduced frontal area \cite{vogel1984,alben2002,gosselin2010,de2012,schouveiler2006,schouveiler2013}. This reconfiguration reduces the load compared to that of a non-deformable body; nonetheless, drag still monotonically increases with speed, although at a lower rate. However, there are a few instances where drag has been observed to drop sharply with velocity. One such phenomenon occurs due to a sudden change in flow regime, referred to as drag crisis, which happens when the boundary layer transitions from laminar to turbulent, moving its separation downstream and narrowing the wake \cite{wieselsberger1921,achenbach1972}. Another scenario involves abrupt and significant shape changes in the body, which can notably occur due to partial breakage, as observed in wind-induced tree pruning \cite{lopez2014}. While this process effectively provokes a sharp decrease in the drag force and preserves the integrity of the rest of the structure, it is not reversible. 

An attractive alternative is snap-through events in flexible structures, during which the elastic system undergoes a rapid transition from one equilibrium state to another. This process is swift, repeatable and produces significant shape changes, making it increasingly popular in engineering for obtaining switchable optical properties, reusable energy trapping materials, microfluidic pumps, or in soft robotics \cite{han2004,tavakol2014,overvelde2015,shan2015}. Snapping instabilities occur through the application of an external input like a point load, electrostatic forces, or thermal effects, and can also be triggered by fluid flows. For example, a post-buckled sheet snaps-through and snaps-back in response to air flowing at different speeds, thus opening and closing an air inlet for flow regulation applications \cite{arena2017,arena2018}. The snapping of a slender membrane or arc has also been used to modify a channel resistance for passive control of viscous flow \cite{gomez2017,peretz2020}, or for energy harvesting from wind-induced snapping oscillations \cite{kim2020flow,kim2021}. The critical flow speed to initiate the instability can be adjusted through the non-linear structural mechanics of the post-buckled member. Note that the system passively responds to flows, which makes it inherently autonomous and leverages the need for additional complex actuation. A few studies focused on the effect of such fast-changing geometries on aerodynamic loading.  In the context of wind turbines, bistable winglets or airfoil trailing-edges autonomously transition to another stable state that alleviates excessive fluid load \cite{arrieta2012,cavens2021}.
While reported drag reduction is limited to about 15\% \cite{cavens2021}, primarily due to a focus on lift applications, it shows potential for more substantial drag drops with appropriate designs.

A promising framework for manufacturing shape-morphing structures with non-linear mechanical properties is the technique of origami, inspired by the Japanese art of paper folding. A surface is equipped with a set of folds, which act as a linkage mechanism, producing large three-dimensional shape changes along selected kinematic pathways that are prescribed by the folding pattern. Origami has also proven to be a powerful method to tailor targeted mechanical properties, and in particular multistability \cite{li2019}. Some specimens can thus settle in more than one stable state in the absence of external load and produce a rapid snap-through response as they pass from one equilibrium to the other \cite{hanna2014,jianguo2015,yasuda2016,li2019}. It offers predictable and reversible shape transformation, with the possibility to tune the shape and number of designated stable configurations through the fold arrangement, crease torsional stiffness, and assigned rest angle \cite{waitukaitis2015}. Those transformable structures with programmable snap-through mechanisms have the potential to produce a controlled drop in drag at targeted flow conditions.

As a starting point, we demonstrate the potential for drag collapse with the waterbomb base, a single vertex bistable origami mechanism that is one of the fundamental origami units \cite{hanna2014,lechenault2014}. This umbrella-like cell has two stable states and features a rapid snap-through response as it transitions from one to the other. Origami structures have already raised interest for drag control, due to their large morphing capabilities \cite{cozmei2020, zhang2021, marzin2022,zhang2022}. In particular, a previous study on the waterbomb base showed that its extreme reconfiguration tends to reduce drag to the point that it no longer increases with flow speed in the regime of large deformation \cite{marzin2022}. Here, we make use of its bistable nature, which has not been utilized before for drag applications. We show that it produces a discontinuous evolution of drag with flow speed, marked by a sudden and significant drop as the structure snaps through. We systematically vary the geometrical and mechanical properties of the unit and show that it allows us to tune the drag discontinuity and the critical speed and loading at which it occurs. Experiments are captured by a simplified theoretical model that is further used for inverse design, which is finding the structural origami parameters that produce a targeted drag collapse.

\section{Results}

\textbf{Drag collapse through snapping}

\begin{figure*}[h]
\includegraphics[width=0.7\linewidth]{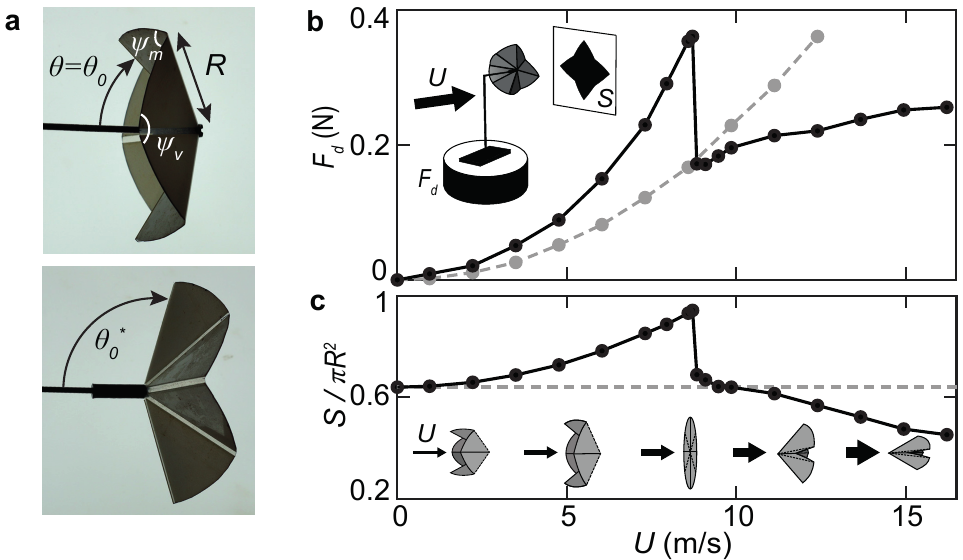}
\centering
\caption{\textbf{Drag collapse through flow-induced snap-through of a bistable origami unit.} (\textbf{a}) Waterbomb base with radius $R$, composed of eight alternating valley and mountain folds with respective angles $\psi_v$ and $\psi_m$, and with unimodal kinematics (for symmetric folding) described by the angle $\theta$. In the absence of loading, the unit sits in one of two stable equilibria with angle $\theta_0$ and $\theta_0^*$. Evolution of (\textbf{b}) the drag force $F_d$ on the bistable origami cell when exposed to a uniform airflow with increasing velocity $U$, and of (\textbf{c}) its dimensionless frontal area $S/\pi R^2$ projected in the plane perpendicular to the flow (see schematics of the experimental set-up in the inset of \textbf{b}). The drag collapse observed in the snapping unit (with $R=5$ cm, $\kappa =19\times 10^{-3}$ N and $\theta_0 = 41^{\circ}$, black solid curves) is contrasted with the quadratic drag experienced by a rigid unit with the same geometry (same $R$ and $\theta_0$, grey dashed curves).}
\label{fig1}
\end{figure*}

The waterbomb base consists of a disk of radius $R$, featuring alternating mountain and valley folds distributed evenly around a central vertex. The present study focuses on the traditional version of the origami waterbomb, comprising a total of eight folds, and which is rigid-foldable. Namely, facets are considered rigid, with all movement occurring at the joints, thus simplifying the kinematic analysis. Additionally, we assume symmetry of motion, based on experimental observation, meaning that all the mountain folds deflect by the same amount, and the same holds for the valley folds. As a result, the mechanism reduces to one degree-of-freedom system \cite{hanna2014,hanna2015}, where the shape of the entire structure can be determined from a single variable, chosen as the angle $\theta$ between a valley fold and the central axis passing through vertex (see Fig.\ref{fig1}a). 

Importantly, this unimodal origami mechanism has two stable equilibrium states and can transition between them through snapping. When at rest, the folded unit sits in one of these stable positions with its vertex pointing in one direction (see Fig.\ref{fig1}a-top). Upon applying an external force, the structure flattens out towards a plane that represents an unstable equilibrium. When displaced beyond this plane, it snaps through and moves to a second stable position with the vertex pointing in the opposite direction (see Fig.\ref{fig1}a-bottom). These two stable positions are characterized by angles $\theta_0$ and $\theta_0^*$, and although they may look similar in Fig.\ref{fig1}a, they are not mirror images of each other. The relationship between the angles of mountain and valley folds, $\Psi_m$ and $\Psi_v$ (measured in terms of deviation of straightness, see Fig.\ref{fig1}a), and $\theta$ also differs on both sides of the unstable equilibrium \cite{hanna2014}. 

The origami bistable mechanism presents the advantage of being readily tunable. The stiffness of the folds notably influences the forces required to transition between stable states. One additional tuning parameter is the rest configuration, characterized by the angle $\theta_0$ in the first stable state. A compact unit (with a smaller $\theta_0$) requires more elastic energy to pass through the unstable planar state, as it has to undergo a greater deflection. The second equilibrium state $\theta_0^*$ also changes correspondingly with $\theta_0$, modulating the amplitude of the shape change upon snapping (the relationship between $\theta_0$ and $\theta_0^*$ is illustrated in Supplementary Fig.5).

Specimens are fabricated using a technique developed in a previous study \cite{marzin2022} and inspired by the literature \cite{overvelde2016,zuliani2018}, which is based on layer superposition. The rigid facets are laser cut from a 350$\mu$m thick sheet of Mylar, which is then attached to a thinner sheet using double-sided tape, forming the creases. These folds act as elastic hinges of specific stiffness and rest angles. Rest angles are prescribed by utilizing the thermo-formable property of PET (see Methods).  As shown in a previous study, the folds then operate as torsional springs, exhibiting elastic behavior around their assigned rest angle \cite{marzin2022}. The stiffness can be adjusted by using different thicknesses of the thin sheet, ranging from $40 -100\mu$m. To measure the stiffness experimentally, a single fold is fabricated using the same process and is tested by tracking the opening angle as an increasing torque is applied to the hinge (see Methods and Supplementary Fig.1). The value of the stiffness is then obtained from the slope of the torque-angle curve. The fabrication process shows a standard deviation of 8$\%$ for $\kappa$.

We conducted tests on ten origami units, sequentially varying the radius $R=4-7$ cm, the stiffness of the folds $\kappa =8-92 \times 10^{-3}$ N, and the rest configuration characterized by the angle $\theta_0=32^{\circ}-66^{\circ}$. These units are subjected to a steady air flow generated by an open jet wind tunnel with a square cross-section of width $40$ cm and flow speed varied in $1-16$ m/s. An elbow arm is used upstream, to hold the unit at its vertex in the center of the cross-section (see inset of Fig.\ref{fig1}b). This arm is connected to a six-component force sensor, measuring the drag component of the fluid force in the direction of the flow. For each measurement, we average the drag over 30 seconds and subtract the drag on the support system alone. The flow-induced shape changes are characterized through the projected area $S$ of the unit in the plane perpendicular to the flow, captured by a camera placed downstream and extracted through Matlab custom image analysis. The observed kinematics align with the unimodal mechanism described earlier, which is rigidly foldable with motion restricted to the joint and exhibits rotational symmetry of motion. The entire geometry of the unit can be reconstructed from $S$, which will be further used as the observable to measure shape reconfiguration (the relationship between $S$ and $\theta$ is illustrated in Supplementary Fig.4).

Fig.\ref{fig1}b-c displays the results of a typical experiment. The unit is initially in its first stable position $\theta_0$, with the concave side facing the flow. As the flow speed gradually increases, the cell symmetrically opens up. The expansion of the frontal area, reported in Fig.\ref{fig1}c, leads to a faster increase of drag with flow speed (solid black curve in Fig.\ref{fig1}b) compared to the classical $U^2$ law observed for a rigid cell frozen in its initial configuration (dashed grey curve in Fig.\ref{fig1}b). When the cell reaches the unstable flat state, it snaps to a more compact and streamlined configuration. Note that slight cell vibrations make it challenging to precisely attain $S/\pi R^2=1$. The sudden reconfiguration results in a discontinuity in the drag force on the unit, and the shape change is significant enough to produce a drag drop by up to $69\%$ among the specimen tested in this study. As $U$ is further increased, the unit closes up, reducing the frontal area exposed to flow. This closing regime has been extensively studied in a previous study \cite{marzin2022}; it features an increase of drag that is slower than the quadratic $U^2$ law for rigid objects, owing to the reduction in frontal area combined with shape streamlining. Variability in the experiments mainly results from the origami fabrication process, yielding a relative dispersion of about $7 \%$ for the drag data and $5 \%$ for the critical snapping velocity (see Supplementary Fig.3).

We thus observe a non-continuous evolution of drag with flow speed with a marked collapse. This behavior can be described by selected observable parameters, namely the critical flow velocity of the snapping $U_c$, the maximum drag force reached just before snapping $F_{d,max}$, and the subsequent drag drop $\Delta F_d$. These parameters are prominent features characterizing the discontinuity but are also relevant for practical applications, such as designing protection devices or valves. They determine a maximum load, the amount of load reduction, or the critical speed at which the structure would operate. In the following section, we investigate how those quantities are related to the origami geometrical and mechanical properties, specifically its size $R$, rest configuration $\theta_0$, and fold stiffness $\kappa$.\\

\textbf{Aeroelastic Mechanism}

\begin{figure*}[h]
\includegraphics[width=\linewidth]{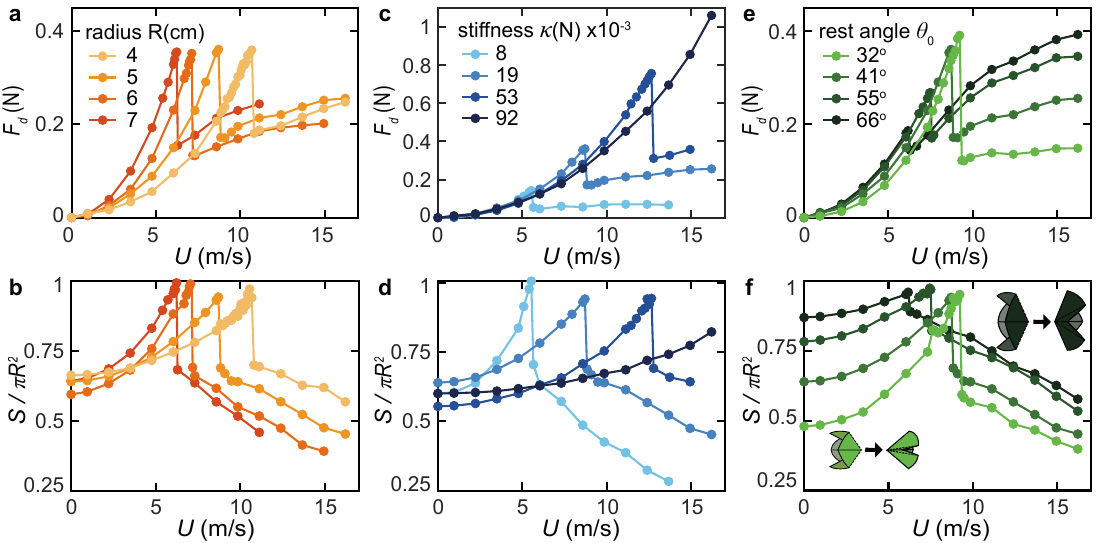}
\caption{\textbf{Influence of the origami structural parameters on drag}. Evolution of (\textbf{a}, \textbf{c}, and \textbf{e}) the drag force $F_d$ and (\textbf{b}, \textbf{d}, and \textbf{f}) the dimensionless frontal area $S/ \pi R^2$ as a function of flow velocity $U$, for origami units with different (\textbf{a}-\textbf{b}) radii $R$, (\textbf{c}-\textbf{d}) fold stiffness $\kappa$, and (\textbf{e}-\textbf{f}) rest angle $\theta_0$. The unit $R=5$ cm, $\kappa=19 \times 10^{-3}N$  and  $\theta_0=41^{\circ}$ is common to all graphs.}
\label{fig2}
\end{figure*}

We first vary the radius $R$ of the unit, while keeping the stiffness to be $\kappa=19\times 10^{-3}$ N and the rest angle to be $\theta_0=41^{\circ}$. Note that fabrication produces a small variability in the rest angle among specimens, which is visible in Fig.\ref{fig2}b in the $6\%$ dispersion in the dimensionless frontal area in the absence of flow ($U=0$). As shown in Fig.\ref{fig2}a-b, all units exhibit the characteristic behavior described earlier, that is a non-continuous evolution of drag with flow speed with a sudden drop associated with the snap-through of the origami unit. However, bigger units (darker orange curves) initially experience a higher fluid loading, because of the larger area exposed to the flow. As a result, they reach the flat state $S/\pi R^2=1$ faster and flip at a lower critical speed (see Fig.\ref{fig2}b). Interestingly, in Fig.\ref{fig2}a, the maximum drag $F_{d,max}$ experienced by the units before snapping is the same, and they feature similar jumps amplitude $\Delta F_d$. Changing the radius of the unit thus primarily shifts the critical velocity $U_c$ at which the drag collapse occurs, while preserving $F_{d,max}$ and $\Delta F_d$.

Fig.\ref{fig2}c-d shows the results for origami cells with different fold stiffness, but the same size $R=5$ cm and rest angle $\theta_0=41^{\circ}$ (with a similar degree of variability as earlier). Stiffer units (darker blue curves) deploy slower with increasing $U$ owing to larger resistance to deformation, and reach the flat state at larger critical velocities (see Fig.\ref{fig2}d). The most rigid one does not reach this unstable point within the range of flow covered here. Snapping at larger speeds $U_c$ results both in higher drag peak values $F_{d,max}$ and larger jumps $\Delta F_d$ (see Fig.\ref{fig2}c). Stiffness thus affects all three drag features, offering an additional lever for control.

Finally, we vary the degree of opening of the cell rest state $\theta_0$, while keeping the size and stiffness constant as $R=5$ cm and $\kappa=19\times 10^{-3}$ N respectively. Starting from a more compact rest state (lighter green curves) delays the flipping towards higher flow velocities, as depicted in Fig.\ref{fig2}f. Intuitively, reaching the flat state that is further away requires greater work from fluid loading. As was the case for the $\kappa$-series of units, snapping at greater $U_c$ results in larger maximal drag $F_{d,max}$ in Fig.\ref{fig2}e. As mentioned earlier and illustrated in the schematics of Fig.\ref{fig2}f, changing $\theta_0$ also affects the degree of closure of the second stable state. Small $\theta_0$ (light green) results in a larger collapse in the frontal area post-snapping compared to units initially more opened (dark green) in Fig.\ref{fig2}f. It correspondingly leads to larger jumps in drag $\Delta F_d$ in Fig.\ref{fig2}e. The rest angle thus completes the set of three parameters that will allow us to adjust the three drag characteristics $U_c$, $F_{d,max}$, and $\Delta F_d$.

\begin{figure*}[h]
\includegraphics[width=0.8\linewidth]{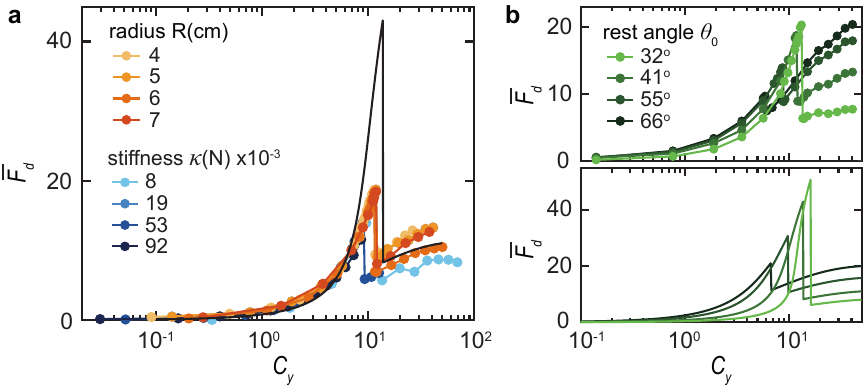}
\centering
\caption{\textbf{Aeroelastic mechanism}. Evolution of the dimensionless drag $\overline{F_d}=F_d/\kappa$ as a function of the Cauchy number $C_y=\rho U^2R^2/\kappa$ for origami units with (\textbf{a}) different radii $R$ and fold stiffness $\kappa$ (while maintaining the same rest angle $\theta_0=41^o$), and (\textbf{b}) different initial degree of opening $\theta_0$ (while keeping the same $R=5$ cm and $\kappa=19 \times 10^{-3}$ N). Experimental data are presented with the same markers and color code as in Fig.\ref{fig2}, and compared to theoretical predictions (solid curves). In panel \textbf{b}, theoretical curves are displayed in a separate graph for readability, and with the same color gradient as in experiments indicating different $\theta_0$. }
\label{fig3}
\end{figure*}

The trends observed through variations in parameters reflect the interplay between external fluid loading and elastic restoring forces of the origami unit, which governs the unit's reconfiguration process. This fluid-elastic competition can be analyzed in terms of Cauchy number $C_y$, as introduced in a previous study \cite{marzin2022} and frequently used in the literature \cite{alben2002,gosselin2010,de2012}. This dimensionless number examines the relative magnitudes of the work done by fluid forces, $\rho U^2 R^3$, and the elastic energy of folding, $\kappa R$:
\begin{equation}
    C_y = \frac{\rho U^2 R^2}{\kappa}
    \label{equation_Cy}
\end{equation}

In Fig.\ref{fig3}a, drag measurements are re-plotted in the dimensionless form $\overline{F_d}=F_d/\kappa$ as a function of $C_y$, using the fold stiffness $\kappa$ as a characteristic force scale, in line with \cite{marzin2022}. Here, we only report data for units with different stiffness $\kappa$ and radius $R$, which share the same initial rest configuration $\theta_0$. Data collapse onto a master curve, with a drag peak and subsequent jump that are solely set by the fold stiffness, and which occur at a critical Cauchy number of  $C_{y,c} \approx 14$ for all units. The dimensionless quantities $C_y$ and $\overline{F_d}$ capture the effect of origami size and stiffness in fluid-elastic mechanisms. However, it does not account for the more intricate influence of the rest angle $\theta_0$. As shown in the top panel of Fig.\ref{fig3}b, the latter impacts the peak and jump value of $\overline{F_d}$, as well as the critical Cauchy number for snapping.

Next, we develop a simplified theoretical model to further explore and corroborate the link between the origami unit's characteristics and drag. This model will also be used for the inverse problem. It employs the same energy approach as previous work on the closing dynamics of the waterbomb unit in a flow \cite{marzin2022} and has been extended to account for the snapping. We provide a brief overview of the main elements of the model here, and additional details can be found in the Methods and in \cite{marzin2022}. The static equilibrium of the unit in a flow is determined through the minimization of its energy $E_{el}-W$, comprising both the elastic potential energy associated with folding $E_{el}$ and the work done by fluid forces $W$ as the unit deforms:

\begin{equation}
\begin{aligned}
E_{el}-W =4\frac{\kappa R}{2}[(\psi_m - \psi_m^0)^2 + (\psi_v - \psi_v^0)^2] \\
- 8 \int_{\theta_0}^{\theta} \int_{S_f}  \rho(\mathbf{U}.\mathbf{n}_{\theta'})^2 dS_f  \mathbf{n}_{\theta'}.d \mathbf{X}_{\theta'}
\label{Eq_en}
\end{aligned}
\end{equation}

Where $\rho$ is the air density and $S_f=\pi R^2/8$ is the facet area (with other quantities introduced subsequently). The first term of Eq.\ref{Eq_en} corresponds to the elastic potential energy of the four mountain and four valley folds, modeled as elastic torsional springs with stiffness $\kappa$. It scales with the square of the deviation of their respective angles $\psi_m$ and $\psi_v$ to their rest values $\psi_{m}^{0}$ and $\psi_{v}^{0}$ \cite{hanna2014}. The second term corresponds to the work done by fluid forces acting on the unit facets, which is the same for all eight facets owing to the rotational symmetry of motion. The Reynolds number for our experiments falls within the range $Re=UR/\nu \approx 10^3 - 10^5$, where form drag predominates over friction drag, and the latter is neglected. As discussed in \cite{marzin2022}, friction forces likely become significant in regimes with substantial deformation, where facets nearly align with the flow. However, this particular regime is not reached in our present study. Consistent with earlier studies on the reconfiguration of flexible structures in a flow \cite{gosselin2010, schouveiler2006, schouveiler2013, marzin2022}, pressure is computed using conservation of momentum principles. It is proportional to the square of the component of flow momentum perpendicular to the facet, denoted as $\rho(\mathbf{U} \cdot \mathbf{n}_{\theta})^2$, where $\mathbf{n}_{\theta}$ represents the normal unit vector. While simplified, this model has demonstrated effectiveness in capturing fluid-elastic behaviors \cite{gosselin2010, schouveiler2006, schouveiler2013, marzin2022}. It offers a reasonable approximation of more complex potential flow expressions \cite{alben2004, marzin2022}, which would be challenging to implement for three-dimensional geometries. The pressure force is integrated over the facet area $S_f$ and its work is computed along the facet trajectory, with $d \mathbf{X}_{\theta}$ the local displacement. By expressing the energy in a dimensionless form as $\overline{E_{el}-W}=(E_{el}-W)/\kappa R$, the previously introduced Cauchy number $C_y$ emerges as the pertinent fluid-elastic parameter:

\begin{equation}
\begin{aligned}
\overline{E_{el}-W} =2[(\psi_m - \psi_m^0)^2 + (\psi_v - \psi_v^0)^2]\\
- 8 C_y \int_{\theta_0}^{\theta} \int_{\overline{S}_f}  \rho(\mathbf{e}_z.\mathbf{n}_{\theta'})^2 d\overline{S}_f  \mathbf{n}_{\theta'}.d \overline{\mathbf{X}}_{\theta'}
\label{Eq_en_adim}
\end{aligned}
\end{equation}

Where $\bar{.}$ denotes dimensionless quantities, using the radius as characteristic length, and $\mathbf{e}_z$ is the unit vector in the direction of the flow. In Eq. \ref{Eq_en_adim}, all kinematic variables of the 1-DOF mechanism depend on a single parameter $\theta$. However, these relationships differ on each side of the flat state. The equilibrium angle $\theta$, which satisfies $d \overline{(E_{el}-W)}/d\theta=0$, is determined numerically while gradually increasing the Cauchy number, first addressing the opening phase of the kinematics. As the system approaches the flat state at $\theta=90^o$ beyond a specific criterion, the kinematic relations are modified to describe the closing phase. From the static equilibrium angle $\theta$, we can determine the drag force acting on the unit as the component of fluid pressure forces in the direction of the flow (see Methods).

Theoretical drag results are presented in Fig.\ref{fig3}a in dimensionless form, considering a given rest angle $\theta_0=41^o$ based on experiments (black solid line). The simplified model successfully captures experimental features, displaying a peak in drag that remains unaffected by the unit's size, followed by a sudden drop at a critical Cauchy number. Although the model tends to overestimate $F_{d,max}$ and $\Delta F_d$, the quantitative agreement is reasonable, considering that no adjustable parameters were used and the simplifications made. As depicted in the lower panel of Fig.\ref{fig3}b, the model also effectively captures the impact of varying the initial rest angle $\theta_0$ on drag. A more compact unit exhibits a higher drag peak and jump, along with a higher critical Cauchy number. This model will be valuable for the inverse design of origami units to achieve the desired drag, as we will discuss in the next section.\\

\textbf{Inverse design approach}
\begin{figure*}[h]
\includegraphics[width=0.9\linewidth]{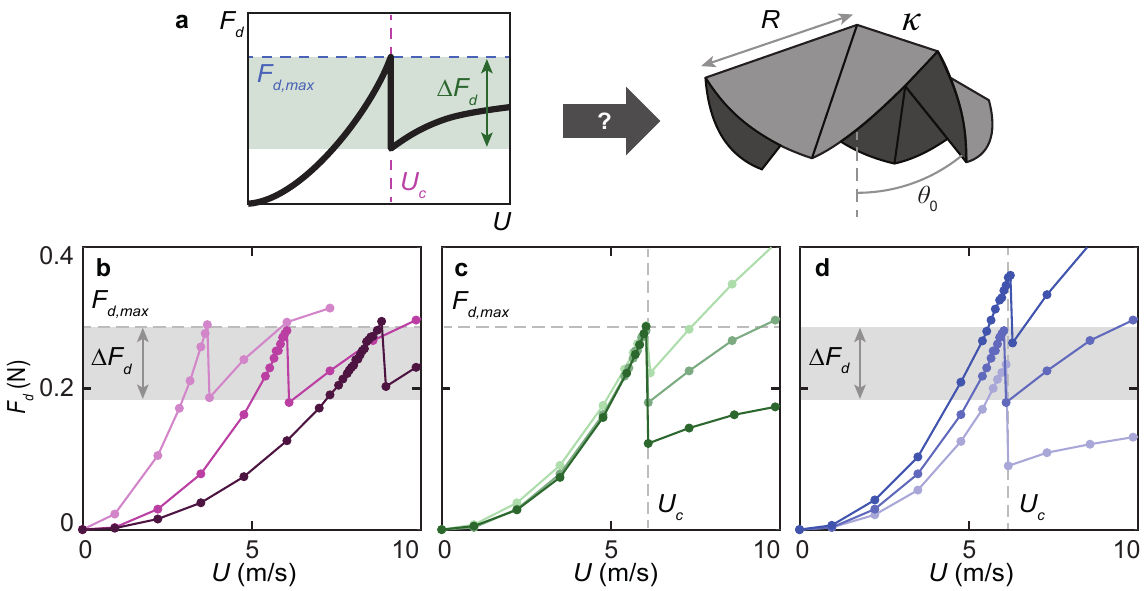}
\centering
\caption{\textbf{Targeting drag collapse features.} (\textbf{a}) The inverse design approach consists of finding the set of origami structural parameters ($R$, $\kappa$, $\theta_0$) that will lead to a collapse of drag with a targeted maximum before snapping $F_{d,max}$, jump $\Delta F_d$, and occurring at a defined critical flow speed $U_c$. Evolution of drag with flow speed for physical prototypes of optimized origami designs, which are calculated to produce a selective variation of (\textbf{b}) $U_c$, (\textbf{c}) $\Delta F_d$, and (\textbf{d}) $F_{d,max}$, while keeping the two remaining features identical. Respective target values of ($U_c$, $\Delta F_d$, $F_{d,max}$) are indicated in the Methods.}
\label{fig4}
\end{figure*}
 
We have characterized how the waterbomb unit features influence its drag behavior. We now pose the question: ``Can we identify the specific geometrical and mechanical origami parameters that would lead to a targeted drag collapse in a uniform fluid flow?" More specifically, we formulate the inverse problem where the objective is to find the optimal set of parameters $(R,\kappa,\theta_0)$ that will result in a collapse of drag, with a predefined target peak drag force $F_{d,max}^t$ and jump $\Delta F_d^t$ value at a critical flow speed $U_c^t$. To address this, we use the model presented earlier and validated experimentally, which establishes a relationship between the parameters $(R, \kappa, \theta_0)$ and the resulting drag characteristics $(F_{d,max}, \Delta F_d, U_c)$. The optimization is implemented numerically in Matlab using the \textit{trust-region-dogleg} algorithm of fsolve solver. It solves the system of equations $F_d-F_{d,max}^t=0$, $\Delta F_d-\Delta F_d^t=0$ and $U_c-U_c^t=0$ with respect to the origami unit's features $(R,\kappa,\theta_0)$. The algorithm finds a single solution, satisfying the constraints by a maximum residual of $O(10^{-19})$ within a hundred iterations. 

To validate the inverse design approach, we test our prediction experimentally by fabricating the optimized design and measuring the evolution of its drag with flow speed. While the specific radius $R$ and rest angle $\theta_0$ can be easily implemented through laser-cutting and an adjustable mount, achieving an arbitrary fold stiffness $\kappa$ is more challenging. This limitation arises from the finite number of sheet thicknesses available for manufacturing the folds. To overcome this constraint, we introduce slits in the folds. These slits modulate the effective length of the folds and, consequently, their stiffness, enabling a controlled variation of $\kappa$ by adjusting their number (see Methods and Supplementary Fig.2).

To demonstrate our ability to control the drag behavior, we create sets of specimens with one parameter gradually changing while keeping the other two fixed. We first vary the targeted critical flow speed for snapping, as $U_c^t=[4, 7, 10]$ m/s, while prescribing constant peak drag value $F_{d,max}^t=0.8$ N and jump $\Delta F_d^t=0.6$ N. The corresponding optimized origami parameters $(R,\kappa,\theta_0)$ are provided in the Methods. Consistent with the differences observed earlier in Fig.\ref{fig3} between experiments and theory, the experimental realizations exhibit lower values in Fig.\ref{fig4}b, namely $U_c=[3.7,6.0,8.8]$ m/s, $F_{d,max}=0.29$ N and $\Delta F_d=0.10$ N. Nevertheless, we successfully achieved a gradual variation in $U_c$ while keeping the other two parameters unchanged. Note that the optimized designs have the same rest angle and fold stiffness, but varying radius, which is consistent with the previous results shown in Fig.\ref{fig2}a. We indeed demonstrated that $F_{d,max}$ and $\Delta F_d$ are size-independent and, therefore, determined by the specific combination of $(\kappa,\theta_0)$. In Fig.\ref{fig4}c and d, we systematically vary the jump $\Delta F_d$ and peak $F_{d,max}$ respectively (target values are reported in the Methods). Similarly, although the physical samples have drag quantities below the target values, they still showcased the intended selective variations. Our results confirm the validity of our inverse design approach, but there are limitations to how closely we can match desired drag behaviors. These limitations arise primarily from the theoretical model employed in the optimization procedure. While it accurately captures the impact of origami parameters on drag, it only provides semi-quantitative agreement with the experimental data. Note that deviations between the targeted drag features and those of the resulting optimized design in Fig.\ref{fig4} align with differences reported in prior experimental-theoretical comparisons of Fig.\ref{fig3} (see Supplementary Fig.7-8). Refining the model, notably using a more realistic representation for the fluid dynamics, is thus expected to produce better quantitative results.

\section{Discussion}

In this study, we leverage the waterbomb base's bistability to sharply reduce drag force with flow velocity. A uniform airflow unfolds the origami unit, which, upon reaching a flat state, undergoes a snap-through to a more streamlined shape, causing a sudden collapse of drag. While the use of elastic deformation to mitigate drag increase with flow speed is not novel \cite{vogel1984,alben2002,gosselin2010,de2012,schouveiler2006,schouveiler2013}, here, a decrease is achieved through significant and abrupt shape changes over a small increment of fluid input load. Importantly, the snap-through is tunable based on the cell radius, fold stiffness, and rest angle, providing control over drag characteristics.\\
We show that the cell behavior results from the quasi-static mechanical equilibrium between actuating fluid loading and restoring elastic forces, captured by a Cauchy number. The snap-through occurs at a fixed critical Cauchy number for a given initial degree of opening. The cell rest angle serves as an additional control, influencing both snap-through onset and drag changes. A compact cell reaches the unstable flat state at a higher $C_y$ and transitions to a correspondingly compact state, producing a larger drop in drag. These experimental features are captured by a theoretical fluid-elastic model, portraying the cell as rigid facets with folds acting as elastic hinges, and utilizing an empirical formulation for fluid pressure forces, consistent with prior studies. The model effectively accounts for the influence of cell structural parameters on drag, demonstrating reasonable predictiveness despite a tendency to overestimate drag. However, such discrepancies are expected, considering the absence of adjustable parameters and the simplified representation of fluid dynamics, which notably does not consider the object’s retroaction on the flow. \\
The modeling framework provides the system equations to guide inverse design, intending to identify the combination of structural parameters leading to a targeted drag collapse. Our implementation showcases an advanced level of control, allowing selective adjustment of key drag features: namely, the peak drag before snapping, the drag drop, and the critical flow speed at which it occurs. Currently, quantitative predictability is contingent on the model’s accuracy. Improved closeness to the target is anticipated with more refined fluid-elastic models, positioning this work as promising proof of concept. It illustrates the potential of origami as a platform for programming drag-vs-speed behavior.\\
This programmability holds value for self-protection strategies, mitigating excessive aerodynamic loads beyond a predetermined threshold. Force transitions above a tunable critical flow velocity could also serve as a velocity threshold detector. The switch-like response of the waterbomb base also holds the potential for fluidic control in a channel, for example, acting as a relief valve. It would obstruct conduct as it deploys and discharges the fluid beyond a predetermined pressure level, limiting pressure built up. In all these applications, morphing is driven aeroelastically, enabling the component to autonomously respond to the local environment, and eliminating the need for additional sensing and control systems. Such passive mechanisms allow for a more streamlined structural design of the aerodynamic control surface.

\begin{figure}[h]
\centering
\includegraphics[width=0.95\linewidth]{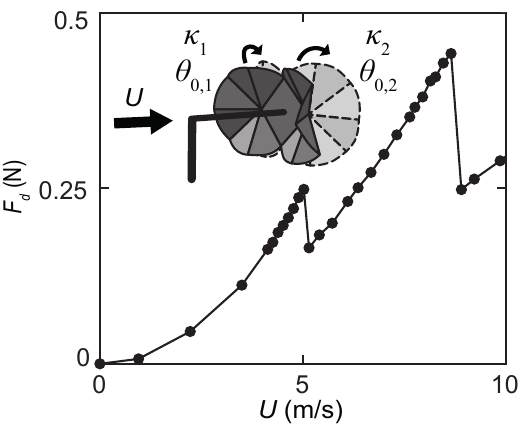}
\caption{\textbf{Leveraging multistability for a double drag collapse.} A multistable origami structure is formed through the assembly of two waterbomb units, with different fold stiffness $\kappa_1=19 \times 10^{-3}$ N and $\kappa_2=53 \times 10^{-3}$ N and rest angle $\theta_{0,1}=56^{\circ}$ and $\theta_{0,2}=71^{\circ}$, and the same radius $R=5$ cm (inset). The successive snap-throughs lead to sequential drag collapses when exposed to a uniform flow with increasing speed.}
\label{fig5}
\end{figure}

An advantage of origami is the ability to easily combine multiple bistable elements to produce multistable structures \cite{silverberg2014,li2015,sengupta2018,jules2022}. An example is shown in Fig.\ref{fig5}, featuring a chain made of two waterbomb cells with the same size and different rest angle and fold stiffness (see also Supplementary Fig.9). It produces two successive sharp transitions in the drag force. With sufficient spatial separation, each unit is expected to have its independent flow velocity threshold initiating snap-through. Other configurations might, however, lead to interactions between cells through external flow, resulting in a more complex collective response, as observed in fluidic cellular origami (albeit for internal pressurization) \cite{li2015}. Alternatively, if cells share facets, they can also communicate the mechanical state of one unit to its neighbor, forming logic elements \cite{treml2018}. This has the potential to generate more complex snapping sequences and subsequent drag evolution with flow speed for advanced adaptive functionalities.

\section*{Methods}

\textbf{Origami fabrication.} Specimens are fabricated using a layer superposition method. The rigid facets are laser cut from a 350 $\mu$m a thick sheet of Mylar, which is then attached to a thinner sheet using double-sided tape with a 2.5 mm spacing that forms the creases. The folded unit is then maintained in a desired configuration on a mount, placed in a heating chamber for an hour at $75^{\circ} C$, and then left to cool down at room temperature. It thus effectively prescribes the first stable state $\theta_0$. This annealing process relaxes any residual stresses that might have been created during folding \cite{lechenault2014} so that the first stable state is close to a zero-energy state.

\textbf{Fold stiffness.} To measure fold stiffness, a prototype of a single crease was designed using the manufacturing process outlined above and in \cite{marzin2022}. While keeping one of the facets vertically fixed, we apply a variable increasing point load to the geometrical center of the free facet and extract the opening angle of the fold $\Psi$ (see Supplementary Fig.1a). The opening angle in the absence of load is set to $\Psi_0 \approx 113^{\circ} \pm 8^{\circ} $ for all the tests, but note that the stiffness was found independent of $\Psi_0$ in the previous study of \cite{marzin2022}. The fold acts as a torsional spring, exhibiting a linear relationship between the torque applied $T$ (accounting for the weight of the hanging facet as well) and the angular deviation $\Psi-\Psi_0$ (see Supplementary Fig.1b). The stiffness is extracted from the slope as $T=\kappa L (\Psi-\Psi_0)$, with $L=4$cm the crease length. Table \ref{tab:stiffness} displays $\kappa$ values for the four thin sheets. Repeatability and robustness were assessed by measuring $\kappa$ for eight folds from a 50 $\mu$m sheet, resulting in an $8\%$ variability based on standard deviation relative to the mean value.
\begin{table}[h]
    \centering
    \begin{tabular}{|c|c|c|c|c|}
    \hline
      Sheet Thickness ($\mu m$)    &  $48$  & $50$  &$75$  & $100$ \\
      \hline
      Fold stiffness $\kappa$(N) $\times 10^{-3}$  &$8 $&$19 $& $53 $ & $92 $ \\
\hline
    \end{tabular}
    \caption{Stiffness measured for folds made from sheets with different thickness. }
    \label{tab:stiffness}
\end{table}

To expand the range of achievable stiffness values, we employ a strategy inspired by prior work \cite{zuliani2018} and introduce slits in the folds. Slits with $5$ mm-length are evenly distributed along the crease length, and span its width (see Supplementary Fig.2a). Adjusting their number modulates the effective length of the fold, thereby changing its stiffness and facilitating the attainment of diverse $\kappa$ values. As shown in Supplementary Fig.2b, the stiffness of a slitted fold scales linearly with its effective length, with a $7 \%$ deviation from the expected linear relationship. To ensure that the presence of slits has minimal impact on the interaction of the origami unit with the flow, we conducted experiments with two specimens —one with slitted folds and another with unslitted folds— both having the same stiffness. These two specimens exhibited identical drag curves, indicating that the slits did not significantly affect the performance in the flow (see Supplementary Fig.2).

\textbf{Fluid-elastic model.} The equilibrium angle $\theta$ of the unit in the flow is given by the zeros of $d\overline{(E_{el}-W)/d\theta}$, with $\overline{E_{el}-W}$ expressed in Eq.\ref{Eq_en_adim}. As further detailed in the Supplementary information, it yields the equation:
\begin{eqnarray}
(\psi_m - \psi_m^0) \frac{d\psi_m}{d\theta} +(\psi_v - \psi_v^0)\frac{d\psi_v}{d\theta} \nonumber\\
 +\frac{2}{3}C_y \sin^2\theta\sin^2\phi \left[A\frac{d\phi}{d\theta} \sin\theta+B\sin\phi \right] =0
 \label{eq_final_theta}
\end{eqnarray}
Where $C_y=\rho U^2 R^2/\kappa$ is the same Cauchy number as defined for experiments, $\phi$ is the angle between valley folds and the central axis, and $A$ and $B$ are constants. Assuming rigid folding and symmetry of motion, the expression for all angles in Eq.\ref{eq_final_theta} can be derived as a function on $\theta$ (see \cite{brunck2016,marzin2022} and Supplementary Information):
\begin{equation}
\phi(\theta)=
\begin{cases}
\arccos \left(\dfrac{\cos\alpha \cos\theta }{1+\sin\alpha\sin\theta}\right) \text{  for } 0 \leq \theta \leq \pi/2\\
\arccos \left(\dfrac{\cos\alpha \cos\theta }{1-\sin\alpha\sin\theta}\right) \text{  for } \pi/2 < \theta \leq \pi - \alpha\\
\end{cases}
\label{eq_phi}
\end{equation}
with $\alpha=2\pi/8$ the sector angle of a facet. Angles of the mountain and valley folds are given by $\cos{\psi_v} = \cos{2\theta}$ and $\cos{\psi_m} = \cos{2\phi}$. Note that the expressions as a function of $\theta$ differ on each side of the flat configuration $\theta=\pi/2$, describing, respectively, the opening and closing kinematics.

Eq.\ref{eq_final_theta} is numerically solved using the nonlinear system solver \textit{fsolve} of Matlab. The Cauchy number is varied linearly from 0 to 50 in 500 steps, and the solver utilizes the value of $\theta$ obtained at the previous iteration as a starting point (initializing at $\theta_0$ for $C_y=0$). As $\theta$ approaches the flat state beyond $\pi/2-\epsilon$, with $\epsilon=0.01$ radians, Eq.\ref{eq_final_theta} is then solved using the new set of angle expressions describing the closing kinematics, with the first starting point at $\pi/2+\epsilon$.

From the equilibrium angle $\theta$, we obtain the dimensionless projected surface $S/\pi R^2=\sin\theta \sin\phi$, and the dimensionless drag force $\overline{F_d}= \pi C_y\sin^3\theta\sin^3\phi$ (by projecting fluid pressure forces in the direction of the flow), for comparison with experimental results.

\textbf{Inverse design.}
The target values $(F_{d,max}^t,\Delta F_d^t,U_c^t)$ used, along with the corresponding cell parameters, are presented in Table \ref{tab:inv_design}. When varying $U_c$ (while keeping the other two drag features constant), cells with the same radius but different $\kappa$ and $\theta_0$ are obtained. Interestingly, selectively varying $\Delta F_d$ is achieved by adjusting $\theta_0$ and $\kappa$ while maintaining a constant $R$.

\begin{table}[h]
    \centering
    \begin{tabular}{|p{0.9cm}|p{0.9cm}|p{0.9cm}||p{0.9cm}|p{0.9cm}|p{0.9cm}|}
    \hline
     \multicolumn{3}{|c||}{\textbf{Target values}} & \multicolumn{3}{c|}{\textbf{Cell parameters}} \\
    \hline
      $F_{d,max}^t$ (N)    & $U_c^t$ (m/s) & $\Delta F_d^t $ (N)& $\kappa$ (N) $\times 10^{-3}$ & $R$ (cm) & $\theta_0 (^\circ)$ \\
      \hline
      0.8 & 4& 0.6 & 21 & 4.6 & 47 \\
      0.8 & 7& 0.6 & 21 & 6.5 & 47 \\
      0.8 & 10 & 0.6 & 21 & 11.4 & 47 \\
      0.8 & 7 & 0.5 & 27 & 6.5 & 56 \\
      0.8 & 7 & 0.7  & 16 & 6.5 & 33 \\
      0.65 & 7 & 0.6 & 11 & 5.9 & 25\\
      0.95 & 7 & 0.6 & 32 & 7.1 & 56 \\      
\hline
    \end{tabular}
    \caption{Fabrication parameters produced by the optimization algorithm, for given target drag parameters.}
    \label{tab:inv_design}
\end{table}


\section*{Data availability}
The datasets generated during the current study are available from the corresponding author upon request.

\section*{Acknowledgments}
We acknowledge support from a JCJC grant of the Agence Nationale de la Recherche (ANR-20-CE30-0009-01) to S.R. and from the Agence Innovation Défense to T.M.

\section*{Author contributions}
S.R., R.N. and T.M. designed the study; R.N. performed the experiments and analyzed the results; R.N. and T.M. performed the theoretical analysis; all authors contributed to interpreting the results and writing the paper.

\section*{Competing interests}
The authors declare no competing interests.

\end{document}


\maketitle
\section{Experimental methods}
\subsection{Fold stiffness}
\begin{figure}[h]
    \includegraphics[width=0.8\linewidth]{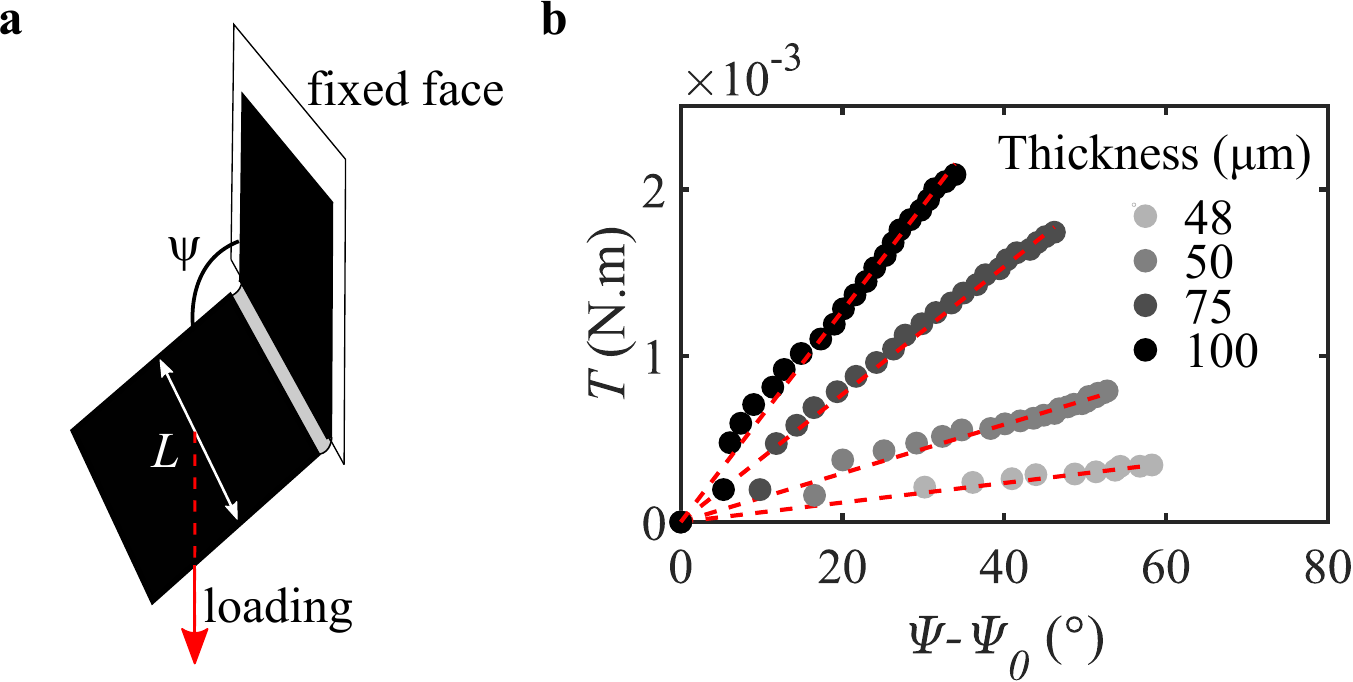}
    \centering
    \caption{(a) Fold stiffness is assessed by measuring the opening angle $\Psi$ of a single fold while applying a variable point force at the center of the free facet. (b) Torque $T$ as a function of the angular deflection relative to the rest angle $\Psi - \Psi_0$, for folds made from sheets of varying thickness.}
    \label{fig:figure2_1}
\end{figure}
To measure fold stiffness, a prototype of a single crease was designed using the manufacturing process outlined in the main text and in \cite{marzin2022}. Specifically, facets are laser-cut in a 350 $\mu$m thick sheet of Mylar and adhered to a thinner sheet using a double-sided tape with a 2.5 mm spacing, constituting the flexural hinge. While keeping one of the facets vertically fixed, we apply a variable point load to the geometrical center of the free facet and extract the opening angle of the fold $\Psi$ (see Fig.\ref{fig:figure2_1}a). The opening angle in the absence of load is set to $\Psi_0 \approx 113^{\circ} \pm 8^{\circ}$ for all the tests, but note that the stiffness was found independent of $\Psi_0$ in the previous study of \cite{marzin2022}. The fold acts as a torsional spring between two rigid facets, exhibiting a linear relationship between the torque applied $T$ (accounting for the weight of the hanging facet as well) and the angular deviation $\Psi-\Psi_0$ (see Fig.\ref{fig:figure2_1}b). The stiffness is extracted from the slope as $T=\kappa L (\Psi-\Psi_0)$, with $L=4$ cm the crease length. Table \ref{tab:stiffness} displays $\kappa$ values for the four thin sheets. As the different mylar sheets were purchased from different companies, we did not investigate stiffness dependence on the thickness.  Repeatability and robustness were assessed by measuring $\kappa$ for eight folds from a 50 $\mu$m sheet, resulting in an $8\%$ variability based on standard deviation relative to the mean value.

\begin{table}[h]
    \centering
    \begin{tabular}{|c|c|c|c|c|}
    \hline
      Sheet Thickness ($\mu m$)    &  $48$  & $50$  &$75$  & $100$ \\
      \hline
      Fold stiffness $\kappa$(N) $\times 10^{-3}$  &$8 $&$19 $& $53 $ & $92 $ \\
\hline
    \end{tabular}
    \caption{Stiffness measured for folds made from sheets with different thickness. }
    \label{tab:stiffness}
\end{table}

\begin{figure}[h]
    \includegraphics[width=\linewidth]{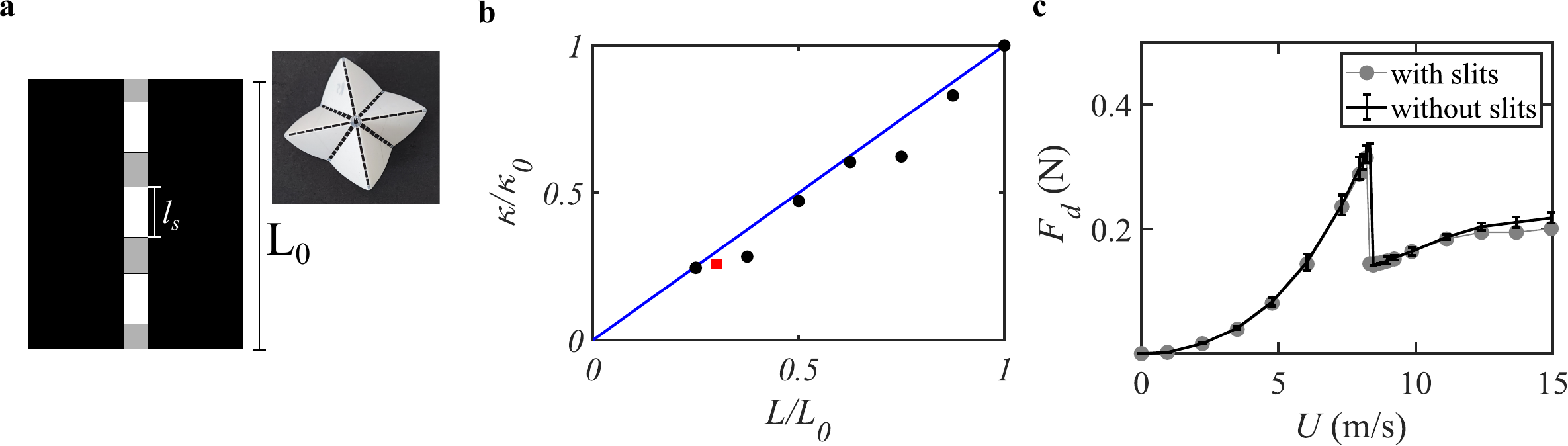}
    \centering
    \caption{(a) To adjust fold stiffness, slits of varying length $l_s$ and number $n_s$, spaced at $5$ mm intervals, are incorporated into the crease. (b) Stiffness $\kappa$ of a slitted fold, relative to the uncut fold stiffness $\kappa_0$, as a function of the ratio of the remaining material length $L=l_s n_s$ to the total length $L_0$, for a fold made from a 75 $\mu$m sheet. The red point corresponds to a stiffness that mimics an uncut fold made from a 50 $\mu$m sheet. (c) Evolution of the drag force as a function of the flow velocity for two origami units with the same fold stiffness: one with slitted folds (grey curve, corresponding to the red point in (b)) and one with unslitted folds (black curve, same curve as in Fig.\ref{fig:figure3}a). For both specimens, $R=5$ cm and $\theta_0=41^o$. }
    \label{fig:figure2_2}
\end{figure}

The finite number of sheet thicknesses available for fold manufacturing imposes limitations on achievable stiffness values, particularly in the context of the inverse design process requiring arbitrary $\kappa$. To address this constraint, we employ a strategy inspired by prior work \cite{zuliani2018} and introduce slits in the folds. Slits with length $l_s=5$ mm are evenly distributed along the crease length $L_0$, and span its width (see Fig.\ref{fig:figure2_2}a). Adjusting their number $n_s$ modulates the effective length of the fold $L=L_0-n_sl_s$, thereby changing its stiffness and facilitating the attainment of diverse $\kappa$ values. As shown in Fig.\ref{fig:figure2_2}b, the stiffness of a slitted fold scales linearly with its effective length, with a $7\%$ deviation from the expected linear relationship.

To verify the minimal impact of slits on the interaction of the origami unit with the flow, we conducted experiments with two specimens —one with slitted folds and another with unslitted folds— both having the same stiffness. The stiffness of folds made from a 50 $\mu$m thick sheet is replicated using a 75 $\mu$m thick sheet with slits corresponding to the red point in Fig.\ref{fig:figure2_2}b. As shown in Fig.\ref{fig:figure2_2}c, these two specimens exhibit nearly identical drag curves, with only a slight deviation at higher flow velocities. This similarity indicates that the slits have minimal impact on the unit's performance in the flow.

\subsection{Repeatability of the experiments}
\begin{figure}
    \includegraphics[width=0.8\linewidth]{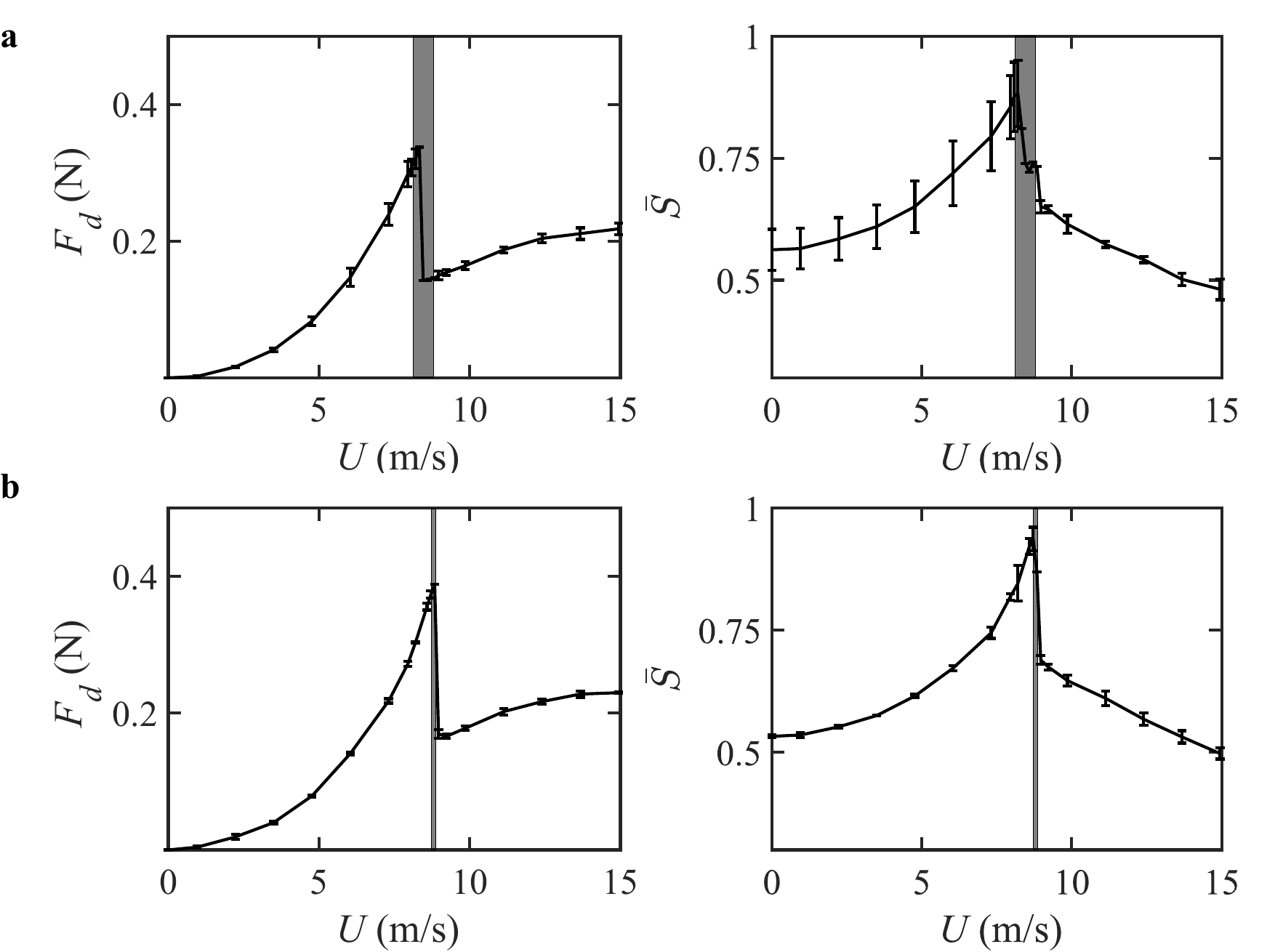}
    \centering
    \caption{ Evolution of drag force $F_d$ and dimensionless projected area $\overline{S}=S/\pi R^2$ with flow speed $U$, for origami specimens with $R=5$ cm, $\kappa = 19 \times 10^{-3}$ N and $\theta _{0}=41^{\circ}$. In (a), experiments are replicated across five independent cells sharing the same parameters. In (b), the same cell is tested three times, with re-annealing between each iteration. Data points represent the mean values, and error bars indicate the standard deviation (the grey zone denotes the variation in critical speed).}
    \label{fig:figure3}
\end{figure}

To test the reproducibility of the experiments and of the fabrication process, we repeated experiments in the wind tunnel for five distinct units fabricated with identical parameters $R=5$ cm, $\kappa = 19 \times 10^{-3}$ N and $\theta _{0}=41^{\circ}$ (which is the common configuration for all three parametric series of origami cells). Fig.\ref{fig:figure3}a presents the mean values for drag and the cell projected area, with error bars corresponding to the standard deviation. The grey region denotes the variation in the critical speed for snap-through $U_c$. The associated relative errors are of the order of $7\%$, $9.5\%$ and $5\%$, respectively for $F_d$, $S/\pi R^2$ and $U_c$. The higher spread in the projected area is attributed to the use of instantaneously captured images, while the units tend to slightly flutter when approaching the unstable planar state.

To pinpoint the primary source of error, we also conducted three repetitions of the experiments using the same unit. Before each test, the unit underwent an hour of re-annealing in the oven, followed by an additional hour of cooling at room temperature to ensure a consistent initial cell configuration. The mean values and standard deviations for the measurements are depicted in Fig.\ref{fig:figure3}b.  The relative error in both plots is within $2\%$. The reduced variability compared to Fig.\ref{fig:figure3}a suggests that the predominant source of variability comes from the fabrication process rather than the experimental measurement method, notably concerning the detection of the snap-through threshold.

\section{Kinematics and mechanical properties of the waterbomb base}
\subsection{ Folding kinematics}

\begin{figure}[h]
    \includegraphics[width=0.9\linewidth]{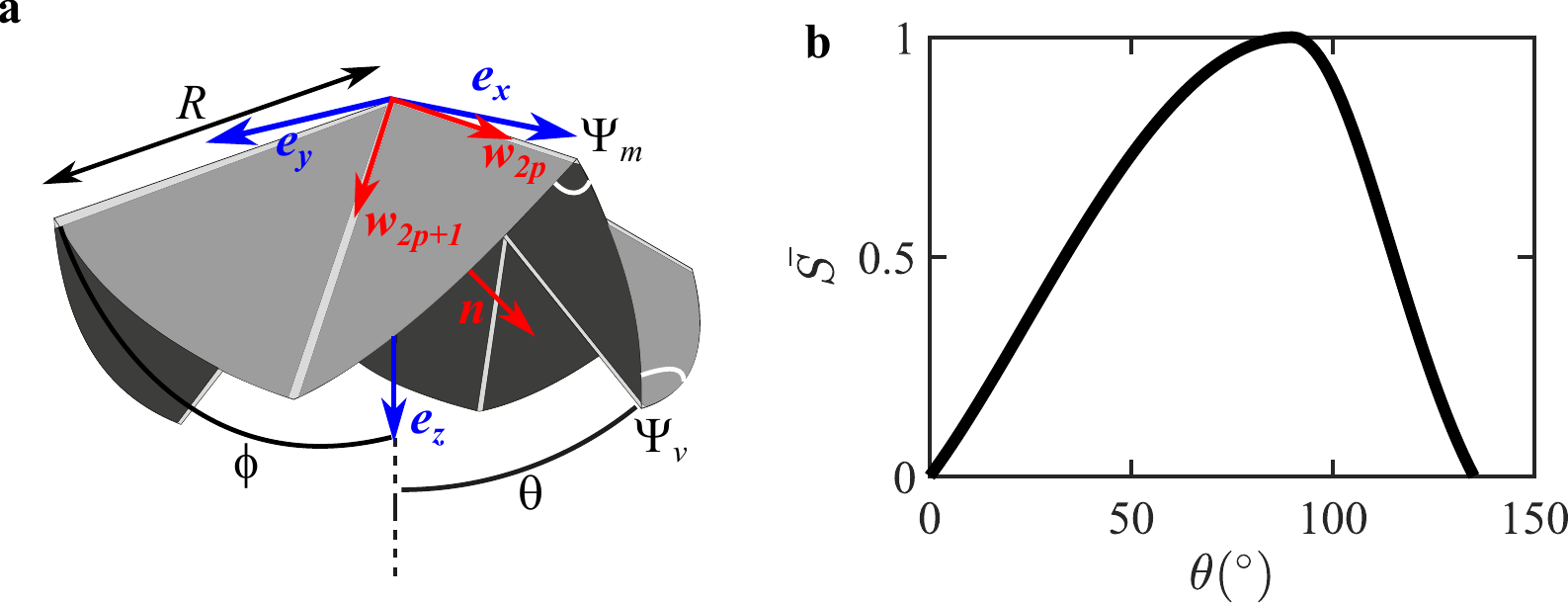}
    \centering
    \caption{(a) Waterbomb base, featuring the various relevant angles (see text), and the unit vectors $\bm{w_{2p}}$  and $\bm{w_{2p+1}}$ aligned along the mountain and valley creases. (b) Variation of the dimensionless projected area $\overline{S}=S/\pi R^2$ with the opening angle $\theta$.}
    \label{fig:figure4}
\end{figure}

The folding kinematics of the waterbomb base have been described extensively in previous work \cite{hanna2014,brunck2016,marzin2022}, and we only recapitulate here the main elements. The traditional version comprises a disk with a radius $R$, featuring $N=8$ folds that radiate from the center and are separated by equal sector angles $\alpha=2 \pi/N$. We assume that the facets are rigid with deformation localized at the creases and that the deformation is symmetric, meaning that all the mountain folds deflect by the same amount, and the same holds for the valley folds. Folding then reduces to one degree of freedom mechanism, which can be described by a single parameter, selected here as the angle $\theta$ made by a valley fold with the central axis passing through the vertex of the unit (see Fig.\ref{fig:figure4}a).

In line with \cite{brunck2016,marzin2022}, the folding kinematics is derived by using the unit vectors $\bm{w_{2p}}$ and $\bm{w_{2p+1}}$, which are aligned respectively with the mountain and valley creases, with $p=0,...,3$ and $\bm{w_0}=\bm{w_8}$. The components of these vectors in the Cartesian system coordinate, shown in Fig.\ref{fig:figure4}a, are:

\begin{equation}
\begin{aligned}
\bm{w_{2p}}=(\sin\phi \cos2p\alpha,\sin\phi \sin2p\alpha,\cos\phi)\\
\bm{w_{2p+1}}=(\sin\theta\cos(2p+1)\alpha,\sin\theta\sin(2p+1)\alpha,\cos\theta)\\
\end{aligned}
\end{equation}

\noindent with $\phi$ the angle between valley folds and the central $z$-axis (see Fig.\ref{fig:figure4}a).  The rigid facet assumption, expressed as $\bm{w_{2p}}.\bm{w_{2p+1}}=\cos \alpha$, establishes a relationship between $\theta $ and $\phi$ :
\begin{equation}
\cos\phi \cos\theta - \cos\alpha(1 - \sin(\phi)\sin(\theta))=0
\label{eq_rig_facet}
\end{equation}

Equation \ref{eq_rig_facet} can be rewritten as :
\begin{equation}
\phi(\theta)=
\begin{cases}
\arccos \left(\dfrac{\cos\alpha \cos\theta }{1+\sin\alpha\sin\theta}\right) \text{  for } 0 \leq \theta \leq \pi/2\\
\arccos \left(\dfrac{\cos\alpha \cos\theta }{1-\sin\alpha\sin\theta}\right) \text{  for } \pi/2 < \theta \leq \pi - \alpha\\
\end{cases}
\label{eq_phi}
\end{equation}

The first relation describes the opening kinematics and the second one describes the closing past the flat configuration $\theta=\pi/2$. The angles of the mountain and valley folds denoted as $\psi_m$ and $\psi_v$ respectively, can be expressed in terms of $\theta$ and $\phi$ \cite{brunck2016}:
\begin{equation}
\begin{aligned}
\cos{\psi_v} = \cos{2\theta}  \\
\cos{\psi_m} = \cos{2\phi} 
\end{aligned}
\label{eq_mount_val}
\end{equation}
Their expression as a function of $\theta$ thus differs as well on each side of the flat configuration.

Note that in the main text, we use $S$, the projected area in the $xy$-plane, to characterize the shape reconfiguration of the unit. Its non-dimensional expression (normalizing by the area of the flat state $\pi R^2$) is given by:
\begin{equation}
\overline{S} =\frac{S}{\pi R^2}=\bm{n}.\bm{e_z}=\sin\theta \sin\phi 
\label{eq_S}
\end{equation}
Where $\bm{n}$ is the unit vector normal to the facet defined by vectors $\bm{w_{2p}}$ and $\bm{w_{2p+1}}$:
\begin{equation}
\bm{n}=\frac{\bm{w_{2p}}\times \bm{w_{2p+1}}}{||\bm{w_{2p}}\times \bm{w_{2p+1}}||}
\label{eq_n}
\end{equation}
The evolution of the dimensionless projected area $\overline{S}$ with the opening angle $\theta$ as the unit opens up and closes is shown in Fig.\ref{fig:figure4}b.

\subsection{Elastic potential energy and stable states}

\begin{figure}[h]
    \includegraphics[width=0.8\linewidth]{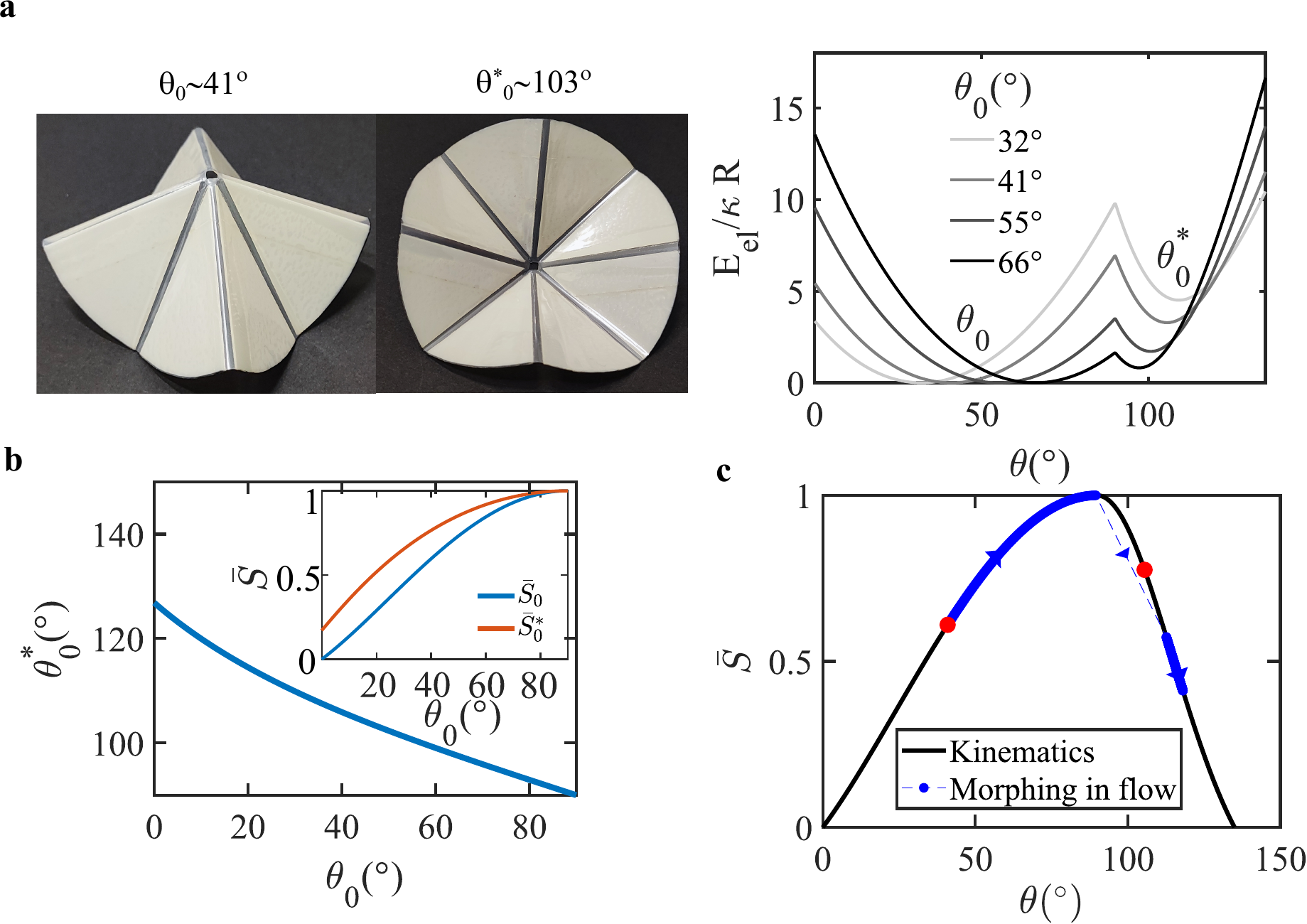}
    \centering
    \caption{(a) Elastic potential energy landscape for the waterbomb base, displaying two stable states with angles $\theta_0$ and $\theta_0^*$, separated by an unstable planar state, for varying $\theta_0$. On the right, images of the two stable states for a unit with $\theta _{\circ}=41^{\circ}$. (b) Evolution of $\theta_0^*$ with $\theta_0$. Corresponding dimensionless projected areas $\overline{S_0}$ and $\overline{S_0^*}$ are shown in the inset. (c) Theoretical prediction for the shape reconfiguration (expressed in terms of $\theta$ and $\overline{S}$) of an origami unit exposed to a flow with increasing velocity (blue curve). It is computed for a waterbomb cell with $\theta_0=41^o$ and varying Cauchy number $C_y=0 - 50$. The black curve represents the waterbomb folding kinematics and the red dots denote the two stable states $\theta_0$ and $\theta_0^*$.  }
    \label{fig:figure5}
\end{figure}

The mechanical behavior of the origami unit is modeled by treating its folds as elastic hinges with rigidity $\kappa$ and length $R$. The elastic potential energy comprises contributions from the four mountain folds and the four valley folds:
\begin{equation}
E_{el}= \frac{4\kappa R}{2} [(\psi_m - \psi_m^0)^2 + (\psi_v - \psi_v^0)^2]
\end{equation}
Here, $\psi_m^0$ and $\psi_v^0$ denote the rest angles of the mountain and valley folds. Using the angle expressions from Eq.\ref{eq_phi} and Eq.\ref{eq_mount_val}, $E_{el}$ can be expressed as a function solely of $\theta$ and $\theta_0$. Fig.\ref{fig:figure5}a shows the evolution of the normalized elastic potential energy $E_{el}/\kappa R$ as a function of $\theta$ for various rest angles $\theta_0$. The energy landscape features two wells corresponding to two stable states, which are separated by an unstable flat state at $\theta=\pi/2$. Pictures of an origami unit in its two stable configurations, characterized by angle $\theta_0$ and $\theta_0^*$, are shown in Fig.\ref{fig:figure5}a. Smaller values of $\theta_0$ (corresponding to compact states, light grey curve) have a higher energy barrier to overcome to transition to the second stable state. The second equilibrium configuration is also correspondingly more compact (higher $\theta_0^*$), resulting in a more pronounced shape change upon snapping. The relationship between $\theta_0^*$ and $\theta_0$ can be numerically determined through energy minimization for a given $\theta_0$, and it is illustrated in Fig.\ref{fig:figure5}b. The inset of Fig.\ref{fig:figure5}b also shows the evolution of the dimensionless projected area $\overline{S_0}$ and $\overline{S_0^*}$ for the two stable states, showing that the two states are not the mirror image of each other and that the second one is always slightly more open than the first one.

 The blue curve in Fig.\ref{fig:figure5}c depicts the morphing of a waterbomb base along its folding pathway (black curve) when subjected to a flow with increasing speed $U$. This evolution is computed using the fluid-elastic model detailed in the next section. In the absence of flow, the unit resides in its first stable state (denoted by the red point). As $U$ rises, it unfolds, and upon reaching the flat unstable state ($\overline{S}=1$), it abruptly snaps towards a state that is more compact than the state $\overline{S}_0^*$ due to the fluid loading.   

\section{Fluid-structure model}
\begin{figure}[h]
    \includegraphics[width=0.4\linewidth]{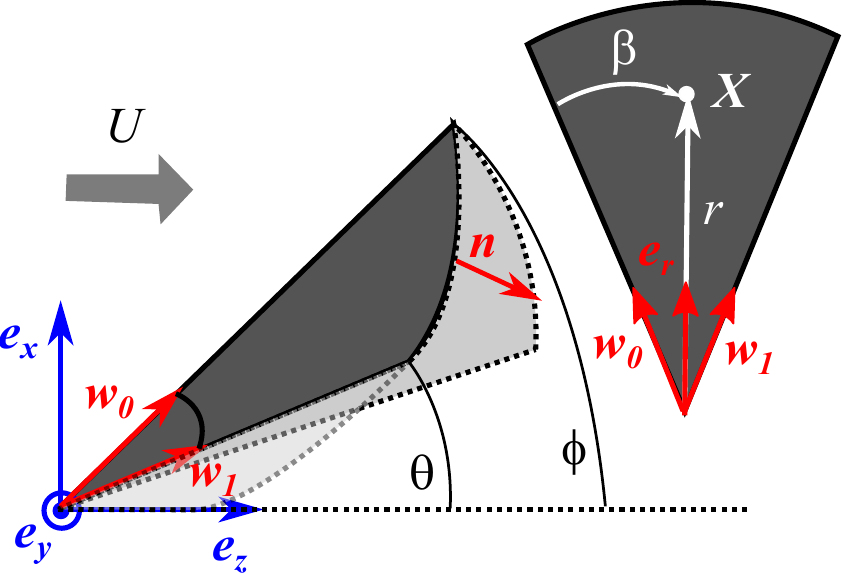}
    \centering
    \caption{Parameterization used to compute the work of fluid forces on a facet defined by the unit vectors $\bm{w_0}$ and $\bm{w_1}$.}
    \label{fig:flow_model}
\end{figure}
To model the origami unit's response to a flow, we adopt the same energy-based approach from prior work on the waterbomb unit's closing dynamics of the waterbomb in a flow \cite{marzin2022}. We recapitulate here the main elements and extend it to include snapping. The unit's static equilibrium in a flow is determined by minimizing its energy $E_{el}-W$, which includes
both the elastic potential energy $E_{el}$, and the work done by fluid forces $W$ as the unit deforms from its initial rest state $\theta_0$ to an angle $\theta$:

\begin{equation}
E_{el}-W =\frac{4\kappa R}{2}[(\psi_m - \psi_m^0)^2 + (\psi_v - \psi_v^0)^2] - 8 \int_{\theta_0}^{\theta} \int_{S_f}  {\rho}(\bm{U}.\bm{n})^2 dS_f  \bm{n}.d\bm{X}
\end{equation}
Where $\rho$ denotes the density of the fluid and other quantities are introduced later. The second term sums the work done by fluid loading on each facet (with area $S_f=\pi R^2/8$), which is the same for all of the eight facets due to the symmetry of motion. Fluid pressure forces acting on each surface element of a facet are integrated along the local displacement $d\bm{X}$. Note that $\bm{n}$ and $\bm{X}$ are functions of the folding angle $\theta$, as indicated by the $\theta$-subscripts in Eq.2 and 3 of the paper. For simplicity, we will omit that subscript here. Pressure is modeled using the momentum conservation arguments, following the approach in \cite{gosselin2010, schouveiler2006,schouveiler2013,marzin2022}. It scales with the momentum carried by the flow in the direction perpendicular to the facet $\rho (\bm{U}.\bm{n})^2$, with $\bm{n}$ the normal unit vector as defined in Eq.\ref{eq_n}.

The trajectory $\bm{X}$ of each surface element of a facet is obtained from the kinematics analysis presented earlier. We consider the facet of Fig.\ref{fig:flow_model} associated with the crease vector field $\bm{w_0}$ and $\bm{w_1}$. A point on the facet is localized through the distance $r \in [0, R]$ and polar angle $\beta \in [0, \alpha]$, as shown in Fig.\ref{fig:flow_model}. Its position vector is expressed as $\bm{ X} = r\bm{e_r}$ with $\bm{e_r}$ the radial unit vector. This vector lies in the $(\bm{w_0},\bm{w_1})$ plane and can thus be expressed as a linear combination of these two vectors:

\begin{equation}
\bm{e_r}=\frac{\bm{w_0} + C\bm{w_1}}{||\bm{w_0} + C\bm{w_1}||}= \frac{\bm{w_0} + C\bm{w_1}}{\sqrt{1 + 2C\cos\alpha+C^2}} \text{  with $C$ verifying  } \bm{e_r}.\bm{w_0}=\cos\beta
\label{eq_er}
\end{equation}
Note that an error was identified in the calculations presented in \cite{marzin2022} (with a missing square root in the denominator in Eq.\ref{eq_er}, and a mistaken interchange of $A$ and $B$ in Eq.\ref{eq_AB}). Importantly, despite these errors, the overall trends and conclusions of the model remain unchanged. In fact, the corrected equations result in theoretical predictions that align more closely with experimental results in \cite{marzin2022}.

From Eq.\ref{eq_er}, the parameter $C$ can be expressed as a function of $\alpha$ and $\beta$:
\begin{equation}
C=\frac{-\cos\alpha(\cos^2\beta-1) + \cos\beta\sqrt{(\cos^2\beta-1)(\cos^2\alpha-1)}}{\cos^2\beta-\cos^2\alpha} 
\end{equation}

Deformation of the origami unit by $d\theta$ results in elementary displacements $d\bm{X}$ for points of the facet : 
\begin{equation}
d\bm{X} =r\frac{d\bm{e_r}}{d\theta} d\theta=\frac{rd\theta}{\sqrt{1 + 2C\cos\alpha+ C^2 }} \left(\frac{d\phi}{d\theta}\cos\phi + C\cos\theta \cos\alpha, C\cos\theta\sin\alpha,-\frac{d\phi}{d\theta}\sin\phi - C\sin\theta \right)
\end{equation}

The work done by fluid forces on all of the eight facets then writes:
\begin{equation}
W = -\frac{8 R^3\rho U^2 }{3} \int_{\theta_0}^{\theta}  \sin^2\theta\sin^2\phi \left[A\frac{d\phi}{d\theta} \sin\theta+B\sin\phi \right] d\theta
\end{equation}
With $A$ and $B$ given by:
\begin{equation}
A = \int_0^{\alpha} \frac{1}{\sqrt{1 + 2C\cos\alpha + C^2 }} d\beta \ \ \text{and} \ \
B =\int_0^{\alpha}  \frac{C}{\sqrt{1 + 2C\cos\alpha + C^2}} d\beta
\label{eq_AB}
\end{equation}

The equilibrium angle $\theta$ of the unit in the flow is then given by the zeros of $d(E_{el}-W)/d\theta$, which yields the equation:
\begin{equation}
(\psi_m - \psi_m^0) \frac{d\psi_m}{d\theta} +(\psi_v - \psi_v^0)\frac{d\psi_v}{d\theta}
 +\frac{2}{3}C_y \sin^2\theta\sin^2\phi \left[A\frac{d\phi}{d\theta} \sin\theta+B\sin\phi \right] =0
 \label{eq_final_theta}
\end{equation}
Where $C_y=\rho U^2 R^2/\kappa$ is the same Cauchy number as defined for experiments. Eq.\ref{eq_final_theta} is numerically solved using the nonlinear system solver \textit{fsolve} of Matlab. The Cauchy number is varied linearly from 0 to 50 in 500 steps, and the solver utilizes the value of $\theta$ obtained at the previous iteration as a starting point (initializing at $\theta_0$ for $C_y=0$). As $\theta$ approaches the flat state beyond $\pi/2-\epsilon$, with $\epsilon=0.01$ radians, Eq.\ref{eq_final_theta} is then solved using the new set of equations describing the angles in the closing part of the kinematics (given in Eq.\ref{eq_phi}) with first starting point at $\pi/2+\epsilon$. 

From the equilibrium angle $\theta$, we obtain the dimensionless projected surface $\overline{S}$ using Eq.\ref{eq_S}, and the dimensionless drag force by projecting pressure fluid forces in the direction of the flow $\bm{e_z}$:
\begin{equation}
\overline{F_d}={F_d}/\kappa = \pi C_y\sin^3\theta\sin^3\phi
\end{equation}

\section{Inverse design}

The objective of the inverse problem is to identify the optimal parameter set
$(R, \kappa, \theta_0)$ that achieves a collapse in drag, targeting a peak drag force $F_{d,max}^t$ and a specific jump value $\Delta F_d^t$ at a critical flow speed $U_c^t$. The optimization uses the fluid-elastic model described earlier, which predicts drag evolution with flow speed based on given cell parameters. Using the  \textit{trust region dogleg} algorithm of fsolve solver from Matlab, the system of equations $F_d-F_{d,max}^t=0$, $\Delta F_d-\Delta F_d^t=0$ and $U_c-U_c^t=0$ is solved for the variables $(R,\kappa,\theta_0)$. The algorithm converges to a single solution meeting the constraints, with a maximum residual of $O(10^{-19})$ within a hundred iterations. Note that the optimization algorithm cannot produce $\theta_0$ below $25^o$, likely due to the simplified representation for fluid forces considered in the model, leading to an inflection point in the $F_d(C_y)$ curve for small $\theta_0$.

\begin{table}[h]
    \centering
    \begin{tabular}{|p{1.5cm}|p{1.5cm}|p{1.5cm}||p{1.5cm}|p{1.5cm}|p{1.5cm}|}
    \hline
     \multicolumn{3}{|c||}{\textbf{Target values}} & \multicolumn{3}{c|}{\textbf{Corresponding cell parameters}} \\
    \hline
      $F_{d,max}^t$ (N)    & $U_c^t$ (m/s) & $\Delta F_d^t $ (N)& $\kappa$ ($\times 10^{-3}$N) & $R$ (cm) & $\theta_0 (^\circ)$ \\
      \hline
      0.8 & 4& 0.6 & 21 & 4.6 & 47 \\
      0.8 & 7& 0.6 & 21 & 6.5 & 47 \\
      0.8 & 10 & 0.6 & 21 & 11.4 & 47 \\
      0.8 & 7 & 0.5 & 27 & 6.5 & 56 \\
      0.8 & 7 & 0.7  & 16 & 6.5 & 33 \\
      0.65 & 7 & 0.6 & 11 & 5.9 & 25\\
      0.95 & 7 & 0.6 & 32 & 7.1 & 56 \\      
\hline
    \end{tabular}
    \caption{Fabrication parameters produced by the optimization algorithm, for given target drag parameters.}
    \label{tab:theoretical curves_optimization}
\end{table}

The target values $(F_{d,max}^t,\Delta F_d^t,U_c^t)$ used, along with the corresponding cell parameters, are presented in Table \ref{tab:theoretical curves_optimization}. When varying $U_c$ (while keeping the other two drag features constant), cells with the same radius but different $\kappa$ and $\theta_0$ are obtained. Interestingly, selectively varying $\Delta F_d$ is achieved by adjusting $\theta_0$ and $\kappa$ while maintaining a constant $R$.

\begin{figure}[h]
    \includegraphics[width=0.9\linewidth]{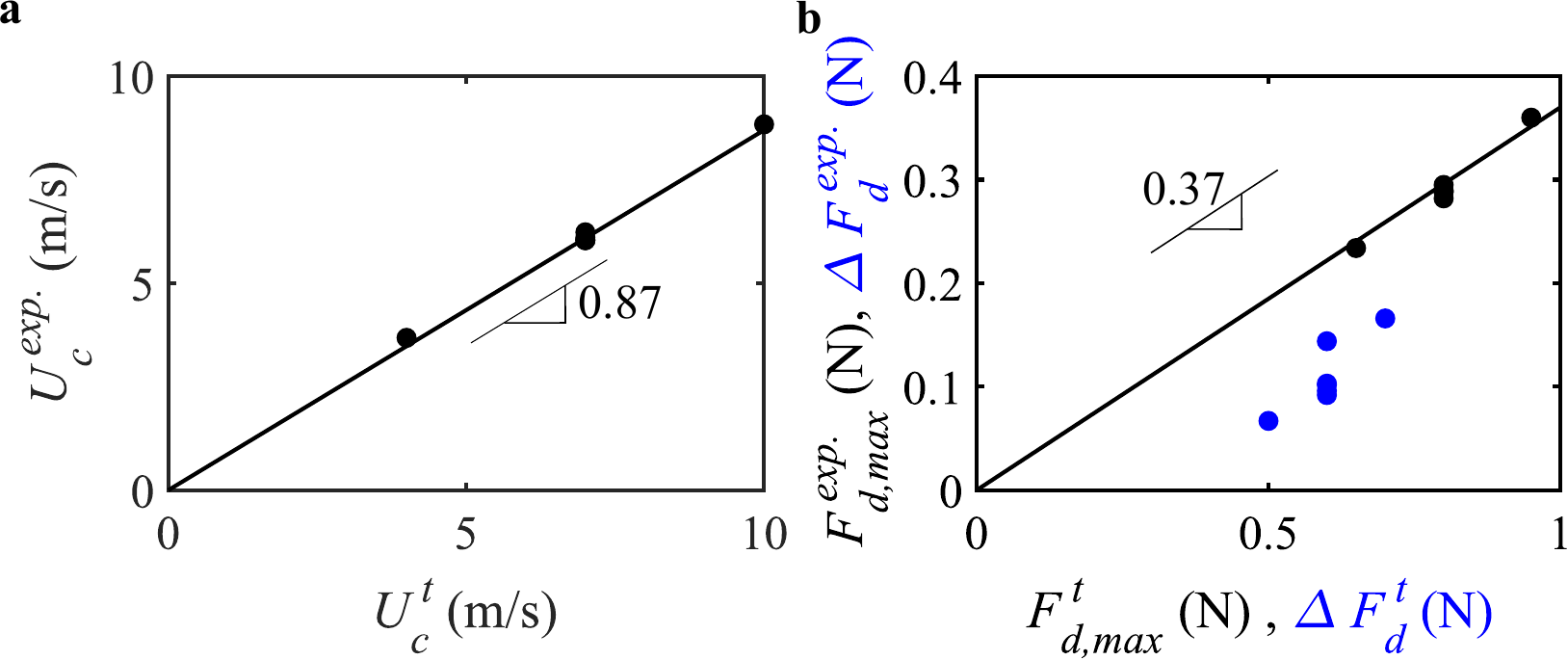}
    \centering
    \caption{Relationship between the target values and the values obtained experimentally for (a) the critical speed $U_c$, and (b) the drag peak $F_{d,max}$ (in black) and jump $\Delta F_d$ (in blue). }
    \label{fig:figure7}
\end{figure}

\begin{figure}[h]
    \includegraphics[width=0.45\linewidth]{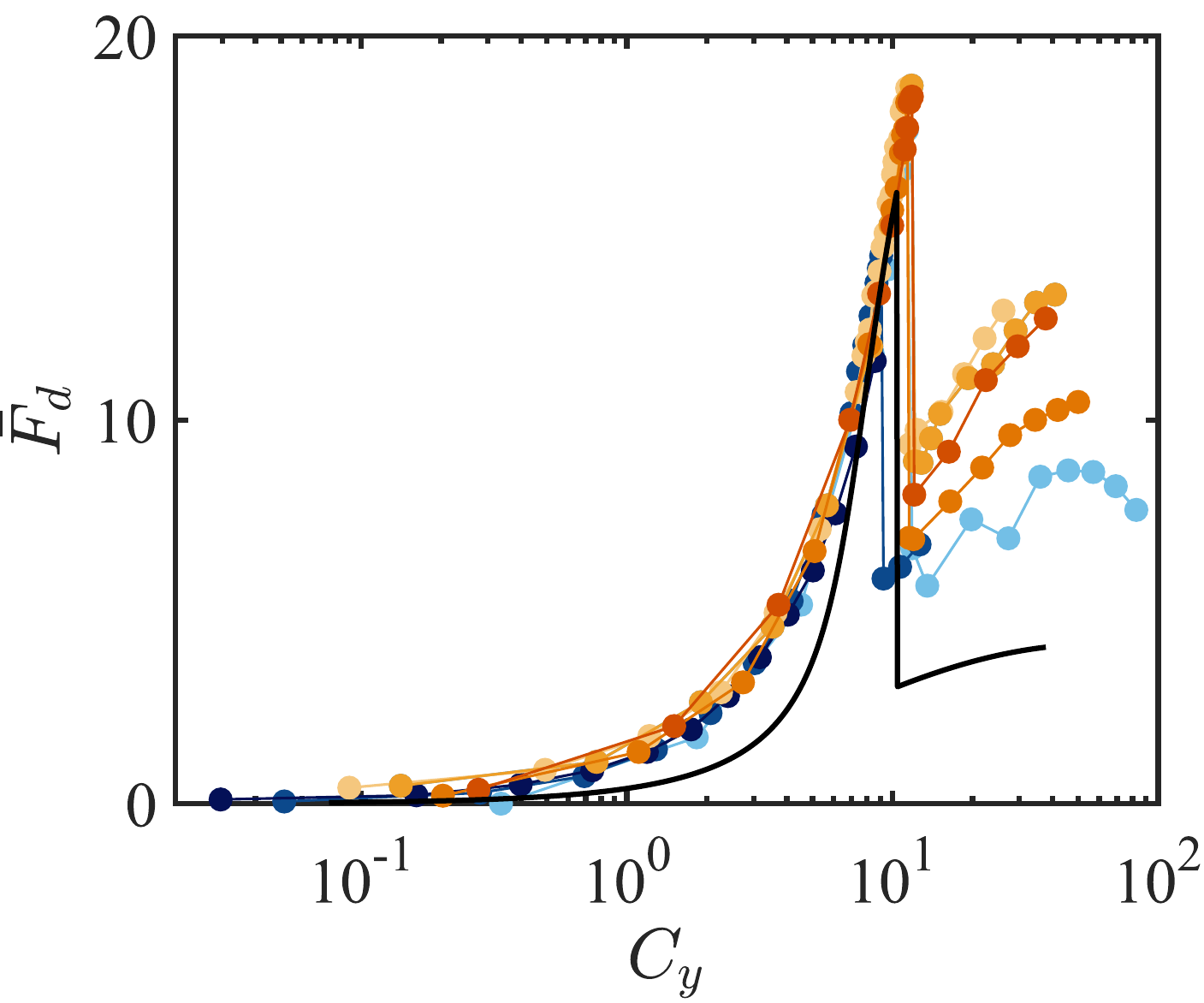}
    \centering
    \caption{Fig.3 of the main text is reproduced with the theoretical curve modified by re-scaling for flow velocity and drag using the proportionality factors from Fig.\ref{fig:figure7}. }
    \label{fig:figure8}
\end{figure}

The physical samples, fabricated using the optimized cell parameters from Table \ref{tab:theoretical curves_optimization}, successfully exhibit the desired selective variation of each drag characteristic (see Fig.4 of the main text). Experimentally obtained values of $(F_{d,max},\Delta F_d,U_c)$ are compared to the target values in Fig.\ref{fig:figure7}. While the experimental values of $U_c$ are reasonably close to the target ones (see Fig.\ref{fig:figure7}a), values of $F_{d,max}$ and $\Delta F_d$ are notably lower than the target values in Fig.\ref{fig:figure7}b. However, this disparity is expected given the semi-quantitative agreement between the model and experiments in Fig.3 of the main text, where the model tends to overestimate drag. From the linear relationships depicted in Fig.\ref{fig:figure7}, we can extract proportionality factors between target values and the ones experimentally obtained, for both flow speed and drag (using $F_{d,max}$ values). In Fig.\ref{fig:figure8}, the model prediction from Fig.3 of the main text is re-plotted by re-scaling drag and flow speed using these proportionality factors, showing a good alignment with the experimental data. Deviations observed in the inverse design implementation are thus consistent with the initial level of agreement between experiment and theory. Using a more refined fluid-elastic model is therefore expected to improve the closeness to the target in the inverse design process. Note that the theoretical $\Delta F_d$ does not align satisfactorily with experimental data in Fig.\ref{fig:figure8}, as the deviation between experimental and target drag value cannot be captured by a single proportionality factor for both $F_{d,max}$ and $\Delta F_d$ in Fig.\ref{fig:figure7}b.

\section{Multiple units}
\begin{figure}[H]
    \includegraphics[width=\linewidth]{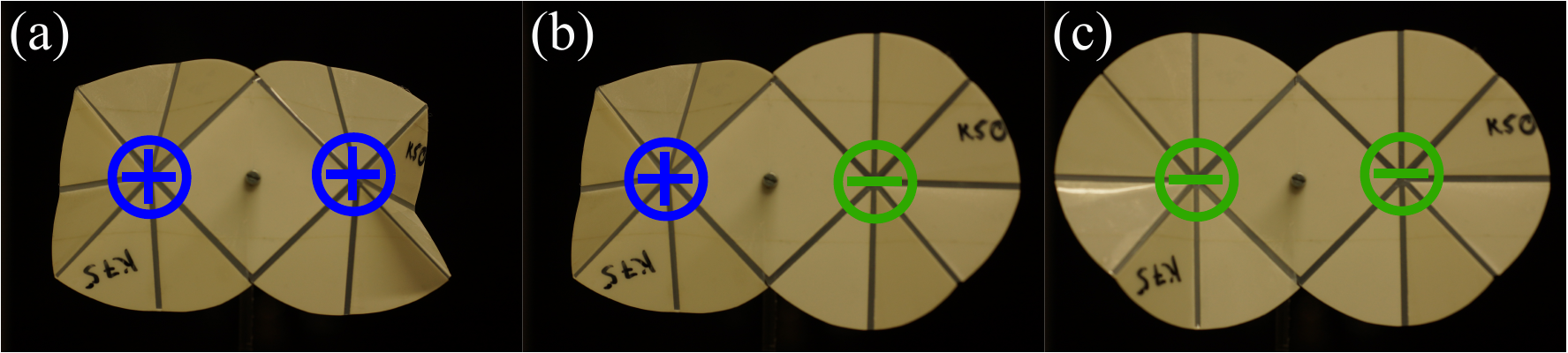}
    \centering
    \caption{Different states of the multistable assembly of the waterbomb units, Frontal view. (a)Both the units are in the first stable state(++). (b)Unit of lower stiffness in a second stable state(-) and higher stiffness in the first stable state(+). (c) Both the units are in their second stable state(--). }
    \label{fig:figure9}
\end{figure}
